\documentstyle[12pt]{article}
\begin{document}
\begin{center}
{\Huge \bf {Acceleration of a Spherical Brane-Universe}
}\\
\vskip 0.3cm
{ Merab GOGBERASHVILI } \\
\vskip 0.3cm {\it
{Andronikashvili Institute of Physics}\\
{6 Tamarashvili St., Tbilisi 0177, Georgia}} \\
\vskip 0.5cm

{\Large \bf Abstract}\\
\quotation {\small The model where the universe is considered as
an expanding spherical 3-brane allows us to explain its expansion
rate without a dark energy component. In this scenario the
computed redshift that corresponds to the transition from cosmic
deceleration to acceleration is in a good agreement with
observations.

}
\endquotation
\end{center}

%%%%%%%%%%%%%%%%%%%%%%%%%%%%%%%%%%%%%%%%%%%%%%%%%%%%%%%%%%%%%%%%%%%%%%%%%%%%%%%

The recent analysis of CMB data combined with other astronomical
observations suggests that the universe is nearly flat, but
possibly with small positive curvature (corresponding to $\Omega =
1,02$), i. e. finite \cite{Sphere}. The simplest space with the
positive curvature is a hyper-sphere. The model with the similar
topology, where the universe is considered as an expanding
spherical 3-brane in 5-dimensional space-time with matter
localized on it, was already introduced by us several years ago
\cite{Gog}. That article even anticipates well-known papers
\cite{Ra-Su} on brane physics. The shell-universe model proposes
obvious explanation of the isotropic runaway of galaxies, the
deficiency in the first modes of the angular power spectrum and
the existence of the cosmological preferred frame. This model also
provides a natural mechanism for the local increment of the brane
tension, leading to the modified Newton's law at galaxy scales,
alternative to galactic dark matter \cite{Go-Ma}.

The recent announcement of the measurement with high-redshift
supernovae indicates a transition from cosmic deceleration to
acceleration \cite{Supernovae}. The accelerated expansion of the
universe usually is model via the cosmological constant. The
problems with introduction of a non-zero cosmological constant in
four dimensions, as well as the relative lack of observational
constraints, inspired many alternative explanations (for recent
reviews see \cite{dark}). A component that causes the accelerate
explanation of the universe is referred to 'dark energy', with the
cosmological constant being just one possibility. In the
shell-universe model, because of the standard cancellation
mechanism \cite{Ra-Su,Gog1}, the cosmological constant on the
brane is zero, while it can be large in the bulk.

Some authors already suggest models without mysterious dark energy
\cite{modified}. In this paper we want to show that if the
universe is considered as an expanding spherical brane in
5-dimensional space-time we can also naturally obtain the
transition from deceleration to acceleration at the observed
redshift without introduction of dark energy on the brane.

We consider Einstein's 5-dimensional equations with negative
cosmological constant
\begin{equation}
R_{AB} - \frac{1}{2}g_{AB}R = - \Lambda g_{AB} + 6\pi^2 G T_{AB}~.
~~~~~~(A, B = 0, 1, 2, 3, 5) \label{Einstein}
\end{equation}
The outer/inner geometries to the shell in the thin-wall formalism
\cite{bubble} are done by \cite{Gog}
\begin{eqnarray}\label{+-}
ds^2_+ = (1 - 2MG/R^2 + \Lambda_+ R^2)dt^2 - (1 - 2MG/R^2 +
\Lambda_+ R^2)^{-1}dR^2 - R^2d\Omega^2 ~,\nonumber\\
ds^2_- = (1 + \Lambda_- R^2)dt^2 - (1 + \Lambda_- R^2)^{-1}dR^2 -
R^2d\Omega^2 ~.
\end{eqnarray}
Here $M$ is the total mass of the spherical universe, $R$ is the
radial coordinate of the hyper-sphere, $G$ is 5-dimensional
Newton's constant, $\Lambda_\pm$ are the outer/inner bulk
cosmological constants, and
\begin{equation}
d\Omega^2 = d\kappa^2 + \sin^2\kappa (d\theta^2 + \sin^2\theta
d\phi^2)
\end{equation}
is the 3-dimensional volume element with the ordinary spherical
angles $\theta$ and $\phi$. The third angle $\kappa$ corresponds
to the ordinary radial coordinate of the Einstein world model
\begin{equation}
dl^2 = \frac{dr^2}{1 - r^2/R^2} + r^2(d\theta^2 + \sin^2 \theta
d\phi^2)
\end{equation}
by the formula $r = R \sin\kappa$.

We suppose that the space-times (\ref{+-}) are separated by a
time-like 3-dimensional thin brane (our universe) with the metric
of the closed isotropic model
\begin{equation} \label{ds4}
ds^2 = d\tau^2  - a^2(\tau ) d\Omega^2 ~,
\end{equation}
where $\tau$ is the intrinsic time of the universe. Let us choose
the equation of a time-like shell in the form:
\begin{equation}
R = a(\tau) ~.
\end{equation}
This means that the cosmological scale factor $a(\tau)$ serves as
the radial coordinate of the 3-shell in our model.

Using ordinary formulae of the thin-wall formalism the equation of
motion of the spherical shell can be written as \cite{bubble}:
\begin{equation} \label{shell}
\sqrt{\dot a^2 + 1 - 2MG/a^2 + \Lambda_+ a^2} - \sqrt{\dot a^2 + 1
+ \Lambda_- a^2} = - \sigma a ~,
\end{equation}
where overdots mean derivatives with respect to $\tau$. In this
equation
\begin{equation}
\sigma  = \frac{1}{3}lim_{\epsilon \rightarrow 0 }
\int\limits_{-\epsilon}^{+\epsilon} (6\pi^2GT^0_0 + \tau^0_0 +
\Lambda) da
\end{equation}
is the intrinsic energy density of the shell-universe, where
$T^0_0$ is matter energy density, $\tau^0_0$ is brane tension and
$\Lambda$ is cosmological constant on the brane. For the
simplicity we can consider the model of matter on the shell with
the equation of state of dust. Since we suppose standard
cancellation of $\Lambda$ and $\tau^0_0$ by the warp factor on the
shell and also that the volume of the 3-sphere in our coordinates
is
\begin{equation}
V = 4\pi a^3 \int_0^\pi d\kappa \sin^2 \kappa = 2 \pi^2 a^3~,
\end{equation}
for $\sigma$ we obtain
\begin{equation} \label{sigma}
\sigma = \frac{GM}{a^3} = 2\pi^2\epsilon G_{(4)}\rho \simeq
\epsilon H^2_0 ~,
\end{equation}
where $G_{(4)}$ is the 4-dimensional gravitational constant, $H_0$
is Hubble's constant, $\epsilon$ is the thickness of the brane and
$\rho = M/2\pi^2R^3$ is the present value of the mean mass density
on the shell.

From (\ref{shell}) one can find the equation for expansion rate of
the shell-universe
\begin{equation} \label{Hubble}
H^2 = \left( \frac{\dot a}{a}\right)^2 = - \frac{1}{a^2} +
\frac{G^2M^2}{\sigma^2a^8} + \frac{(\Lambda_+ - \Lambda_-)^2
}{4\sigma^2} - \frac{\Lambda_+ + \Lambda_-}{2} +
\frac{\sigma^2}{4}+ \frac{GM}{a^4}  - \frac{GM(\Lambda_+ -
\Lambda_-)}{\sigma^2 a^4} ~.
\end{equation}
Using (\ref{sigma}) we notice that second term at right side
exactly cancels the first, curvature, term. Next simplification
may be made from the observation that for the small redshifts $a
>> \epsilon$, using the estimation (\ref{sigma}), we can neglect
the last three terms in (\ref{Hubble}). This means that in
distinct from the 4-dimensional relativity, where expansion rate
of the universe is determined by the matter content and the
curvature, in our case present dynamics of the shell-universe is
governed mainly by the bulk cosmological constants.

Finally, parameterizing the scale factor by the redshift $z$
\begin{equation}
a(\tau) = \frac{R}{1 + z}~,
\end{equation}
where $R$ is the (co-moving) radius of curvature, the equation
(\ref{Hubble}) takes the form:
\begin{equation} \label{Hubbl-z}
H^2 = \left( \frac{\dot z}{1+z}\right)^2 \simeq H^2_0 \left[
\frac{\Omega_-}{(1+z)^6} - \Omega_+ \right]~,
\end{equation}
where
\begin{equation} \label{Omega}
\Omega_- = \frac{(\Lambda_+ -
\Lambda_-)^2}{4\sigma^2H_0^2}~,~~~~~~~~ \Omega_+ = \frac{\Lambda_+
+ \Lambda_-}{2H_0^2}~.
\end{equation}
These parameters give the relative contributions to the present
expansion rate $H_0$ with
\begin{equation}\label{Omega+-}
\Omega_- - \Omega_+ \simeq 1~.
\end{equation}

Using (\ref{Hubbl-z}) we can write also the cosmological
acceleration equation
\begin{equation} \label{acceleration}
 \frac{\ddot a}{a} \simeq H^2_0 \left[
\frac{4\Omega_-}{(1+z)^6} - \Omega_+ \right]~.
\end{equation}
From this equation we see that, because of (\ref{Omega+-}), at the
present stage our universe is in accelerating regime. Since the
constants $\Omega_-$ and $\Omega_+$ have the same order of
magnitude one can find from (\ref{acceleration}) that the
acceleration starts at the redshift
\begin{equation}
z > 4^{1/6} -1 \simeq 0.3~,
\end{equation}
when the first term exceeds the second one. This result is in good
agrement with observational data \cite{Supernovae} and does not
depend much on the particular matter content of the model, but has
geometrical origin.

So it is natural to consider the universe as an expanding
spherical brane with confined matter on it. Some observational
data, such as, isotropic runaway of galaxies, the deficiency in
the first modes of the angular power spectrum and existence of the
preferred frame in the universe, support this model. As it was
shown in this paper, shell-universe model predicts also the
correct value of the redshift parameter that corresponds to the
transition from cosmic deceleration to acceleration.

\medskip
\noindent {\bf Acknowledgements:} The author would like to
acknowledge the hospitality extended during his visits at the
Abdus Salam International Centre for Theoretical Physics where
this work was done.

%%%%%%%%%%%%%%%%%%%%%%%%%%%%%%%%%%%%%%%%%%%%%%%%%%%%%%%%%%%%%%%%%%


\begin{thebibliography}{99}

\bibitem{Sphere} J. L. Tonry,  et al.,
                 {\it Astrophys. J.}  {\bf 594} (2003) 1 [arXiv: astro-ph/0305008]; \\
                 J. -P. Luminet, J. Weeks, A. Riazuelo, R. Lehoucq and
                 J. -P. Uzan J.,
                 {\it Nature} {\bf 425} (2003) 593 [arXiv: astro-ph/0310253].

\bibitem{Gog} M. Gogberashvili,
               {\it Europhys. Lett.} {\bf 49} (2000) 396 [arXiv: hep-ph/9812365].

\bibitem{Ra-Su} L. Randall and R. Sundrum,
                {\it Phys. Rev. Lett.} {\bf 83} (1999) 3370 [arXiv: hep-ph/9905221];
                Ibidem, {\it Phys. Rev. Lett.} {\bf 83} (1999) 4690 [arXiv: hep-th/9906064].

\bibitem{Go-Ma} M.Gogberashvili and M.Maziashvili,
                 {\it Gen. Rel. Grav.} {\bf 37} (2005) 1129 [arXiv: astro-ph/0404117].

\bibitem{Supernovae} R. A. Knop,  et al.,
                     {\it Astrophys. J.} {\bf 598} (2003) 102 [arXiv: astro-ph/0309368]; \\
                     A. G. Riess, et al.,
                     {\it Astrophys. J.} {\bf 607} (2004) 665 [arXiv: astro-ph/0402512].

\bibitem{dark}  P. J. E. Peebles and B. Ratra,
               {\it Rev. Mod. Phys.} {\bf 75} (2003) 559 [arXiv: astro-ph/0207347]; \\
               T. Padmanabhan,
               {\it Phys. Rept.} {\bf 380} (2003) 235 [arXiv: hep-th/0212290];\\
               L. Barnes, M. J. Francis, G. F. Lewis and E. V. Linder,
               [arXiv: astro-ph/0510791].

\bibitem{Gog1} M. Gogberashvili,
               {\it Int. J. Mod. Phys.} {\bf D 11} (2002) 1635 [arXiv: hep-ph/9812296];
               Ibidem, {\it Int. J. Mod. Phys.} {\bf D 11} (2002) 1639 [arXiv: hep-ph/9908347].

\bibitem{modified} C. Deffayet, G. Dvali and G. Gabadadze,
                   {\it Phys. Rev.} {\bf D 65} (2002) 044023 [arXiv: astro-ph/0105068];\\
                   S. Nojiri and S. Odintsov
                   {\it Phys.Rev.} {\bf D 68} (2003) 123512 [arXiv:
                   hep-th/0307288];\\
                   S. M. Carroll, V. Duvvuri, M. Trodden and M. S. Turner,
                   {\it Phys. Rev.} {\bf D 70} (2004) 043528 [arXiv: astro-ph/0306438]; \\
                   S. Capozziello, V. F. Cardone, S. Carloni and A. Troisi,
                   {\it Int. J. Mod. Phys.} {\bf D 12} (2003) 1969 [arXiv: astro-ph/0307018];\\
                   A. Lue, R. Scoccimarro and G. Starkman,
                   {\it Phys. Rev.} {\bf D 69} (2004) 044005 [arXiv: astro-ph/0307034];\\
                   E. W. Kolb, S. Matarrese and A. Riotto,
                   [arXiv: astro-ph/0506534].

\bibitem{bubble} V. A. Berezin, V. A. Kuzmin and I. I. Tkachev,
                 {\it Phys. Rev.} {\bf D 36} (1987) 2919; \\
                 A. Barnaveli and M. Gogberashvili,
                 in {\it New Frontiers in Gravitation} (Hadronic Press, Palm Harbor, 1996)
                 [arXiv: hep-ph/9505412].

\end{thebibliography}
\end{document}